\documentclass[aps,prb,twocolumn,showpacs,floatfix,superscriptaddress,nofootinbib]{revtex4}
\usepackage{graphicx}
\usepackage{hyperref}
\usepackage{amssymb}
\usepackage{amsmath}
\usepackage{lmodern}
\usepackage{color}
\usepackage{empheq}
\usepackage{epstopdf}
\usepackage{amsmath}

\begin{document}

\title{Nonlinear transmission line: shock waves and the simple wave approximation}. 

\author{Eugene Kogan}
\email{Eugene.Kogan@biu.ac.il}
\affiliation{Department of Physics, Bar-Ilan University, Ramat-Gan 52900, Israel}

\begin{abstract}
The transmission lines we consider are constructed from the nonlinear inductors and the nonlinear capacitors. 
In the first part of the paper we additionally include linear ohmic resistors.
The dissipation thus being taken into account leads to the existence of shocks - the travelling waves with the different asymptotically constant values of the voltage (the capacitor charge) and the current before and after the front of the wave. 
For the particular values of ohmic resistances (corresponding to strong dissipation) we  obtain the analytic solution for the profile of a shock wave.
Both the charge
on a capacitor and current through the inductor are obtained as the functions of the time and space coordinate. For the case of weak dissipation we obtain the stationary dispersive shock waves. 
In the second part of the paper we consider nonlinear lossless transmission line.
We formulate the simple wave approximation for such transmission line which decouples left- right-going waves.  The approximation can also be used for the lossy transmission line considered in the first part of the paper to describe the formation of the shock wave (but, of course, not the shock wave itself).

\end{abstract}

\date{\today}

\maketitle

\section{Introduction}

The concept that in a nonlinear wave propagation system
the various parts of the wave travel with different
velocities, and that wave fronts (or tails) can sharpen
into shock waves, is deeply imbedded in the classical
theory of fluid dynamics \cite{whitham}.
The  nonlinear electrical transmission lines 
(where the nonlinearity can be either due to nonlinear capacitors or nonlinear inductors forming the line, or both)
are of much interest both due to their applications, and as the laboratories to study nonlinear waves \cite{malomed2,congy,french,nouri,neto,nikoo,silva,wang,rangel,kyuregyan,akem,fairbanks}.
Nonlinear Transmission Line  technology has historically been used for pulse shaping applications and in digitizing oscilloscopes. Over the years it has proven itself to be a highly credible, robust technology.

A very interesting particular type of signals which can propagate along such
lines - the shock waves - is attracting interest since long ago \cite{landauer1,landauer2}.
We published a series of papers on the travelling waves in nonlinear transmission lines: the kinks, the solitons and the shocks (see the most recent publication of that series and the references therein \cite{kogan6}). 

In the present short note we  would like to add to  our previous publications on the subject. In the first half of the paper we start from reproducing in a concentrated form the analytic results for the profile of the shock wave in the transmission line with the nonlinear capacitors obtained in  Ref. \cite{kogan6}. 
Those results were obtained by reducing the second order ordinary differential equation
describing the travelling waves i the transmission line and factorising the thus obtained equation for the definite values of the parameters. 
While in the latter publication only the charge of the capacitors was analysed as the function of the time and the coordinate (which is equivalent to studying the voltage in the line as the function of the time and the coordinate), in the present note the time and coordinate dependence of the current is also calculated.

The second half of the paper is dedicated to the presentation of the 
simple wave approximation to the wave equation describing general lossless transmission line, which allows  to decouple the nonlinear wave equation into two separate equations for the right- and left-going waves \cite{landau,rabinovich,vinogradova}.
While we used this approximation previously for the Josephson transmission line (JTL) 
\cite{kogan1}, here we formulate the approximation for the case when both the inductors and the capacitors are nonlinear \cite{kogan6}.

\section{The circuit equations}
\label{circ}

\begin{figure}[h]
\includegraphics[width=\columnwidth]{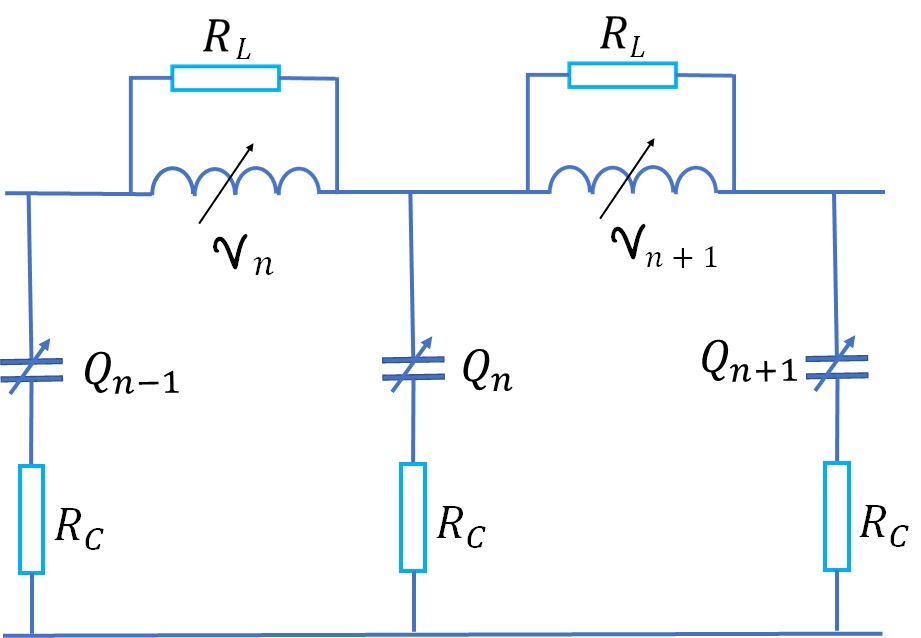}
\caption{Lossy nonlinear  transmission line}
 \label{r5}
\end{figure}

 The   transmission line constructed from the identical nonlinear inductors and the identical nonlinear capacitors is shown on Fig. \ref{r5}.
We take the  capacitors charges $Q_n$ and the integrated voltages on the inductors
\begin{eqnarray}
\Phi_n\equiv\int {\cal V}_ndt
\end{eqnarray}
as the dynamical variables.
The circuit equations (Kirchhoff laws) are
\begin{subequations}
\label{l}
\begin{alignat}{4}
\frac{dQ_n}{dt}&=  I_n-I_{n+1}+\frac{1}{R_L}\frac{d}{dt}\left(\Phi_n-\Phi_{n+1}\right),\label{lb}\\
\frac{d \Phi_{n+1}}{d t}&
=V_{n}-V_{n+1}+R_C\frac{d}{dt}\left(Q_n-Q_{n+1}\right).\label{la}
\end{alignat}
\end{subequations}

To close the system (\ref{l}) we should specify the connection between the voltages on the capacitors $V_n$ and the charges and also between the currents through the inductors $I_n$ and the integrated voltages
\begin{subequations}
\label{q}
\begin{alignat}{4}
V_n&= V(Q_n),\label{qq}\\
I_n&=I(\Phi_n).\label{qq2}
\end{alignat}
\end{subequations}

In the continuum approximation we  treat $n$  as the continuous variable $x$
(we measure distance in the units of the transmission line period)
and approximate the finite differences in the r.h.s. of the equations by the first derivatives with respect to $x$,   after which the equations take the form
\begin{subequations}
\label{c2}
\begin{alignat}{4}
\frac{\partial Q}{\partial t} &=  -\frac{\partial I(\Phi)}{\partial x}
 -\frac{1}{R_L}\frac{\partial^2 \Phi}{\partial x\partial t},\\
\frac{\partial \Phi}{\partial t}&= -\frac{\partial V(Q)}{\partial x}
-R_C\frac{\partial^2 Q}{\partial x\partial t}.
\end{alignat}
\end{subequations}
Further on we'll consider  the  travelling waves, for which all the dependent variables   depend upon the single independent variable $\tau=t-x/U$, where $U$ is the speed of the wave. For such waves (\ref{c2}) turns into the system of ODE
which   after integration takes the form
\begin{subequations}
\label{e}
\begin{alignat}{4}
\frac{1}{R_L}\frac{d \Phi}{d\tau}&=UQ-I(\Phi)+\widetilde{I},\label{eb}\\
R_C\frac{d Q}{d\tau}&=U\Phi-V(Q)+\widetilde{V}, \label{ea}
\end{alignat}
\end{subequations}
where $\widetilde{I}$ and $\widetilde{V}$ are the constants of integration.

Equation (\ref{e}) contains 3 arbitrary constants - $U$,
$\widetilde{I}$ and $\widetilde{V}$. 
Since we are considering localized travelling waves, 
it would be convenient to take as the constants  $Q_1,Q_2,\Phi_1,\Phi_2$, defined by the equations
\begin{subequations}
\label{bc}
\begin{alignat}{4}
\lim_{\tau\to-\infty}Q(\tau)&=Q_1; \hskip .5cm \lim_{\tau\to+\infty}Q(\tau)=Q_2; \label{bca}\\
\lim_{\tau\to-\infty}\Phi(\tau)&=\Phi_1; \hskip .5cm \lim_{\tau\to+\infty}\Phi(\tau)=\Phi_2,
\label{bcb}
\end{alignat}
\end{subequations}
and connected by the relation
\begin{eqnarray}
\frac{Q_2-Q_1}{\Phi_2-\Phi_1} =  \frac{I_2-I_1}{V_2-V_1}.
\end{eqnarray}
Thus Eq. (\ref{e}) can be presented as
\begin{subequations}
\label{er}
\begin{alignat}{4}
\frac{1}{R_L}\frac{d \Phi}{d\tau}&=U(Q-Q_2)-I(\Phi)+I_2,\label{ebr}\\
R_C\frac{d Q}{d\tau}&=U(\Phi-\Phi_2)-V(Q)+V_2, \label{ear}
\end{alignat}
\end{subequations}
where $I_2=I(\Phi_2)$ and $V_2=V(Q_2)$.

The speed of the travelling wave is 
\begin{eqnarray}
U= \frac{I_2-I_1}{Q_2-Q_1}=\frac{V_2-V_1}
{\Phi_2-\Phi_1}.
\end{eqnarray}

Further on we consider transmission line with the linear inductors ($\Phi=LI$).
and nonlinear capacitors.
We can exclude the current from (\ref{er})  and after a bit of algebra obtain
closed equation for $Q(\tau)$
\begin{eqnarray}
\label{199}
\frac{LR_C}{R_L}\frac{d^2Q}{d\tau^2}+R_C\frac{dQ}{d\tau}
+\frac{L}{R_L}\frac{dV(Q)}{d\tau}\nonumber\\
= U^2L(Q-Q_2)-V(Q)+V_2,
\end{eqnarray}
and the travelling speed is given by the equation
\begin{eqnarray}
U^2=\frac{1}{L}\frac{V_2-V_1}{Q_2-Q_1}.
\end{eqnarray}
The solution of (\ref{199}) should satisfy the boundary conditions (\ref{bca}).

Let us   approximate  $V(Q)$  for $Q$ between $Q_1$  and $Q_2$  as
\begin{eqnarray}
\label{we}
V(Q)=\text{const}+\frac{Q+\beta Q^2}{C_s},
\end{eqnarray}
where $\beta$ and $C_s$ are constants.
Let us also take $V(Q)$ in the last term in the l.h.s. of (\ref{199}) in the linear approximation.
In this cases,   (\ref{199}) becomes
\begin{eqnarray}
\label{29}
T^2\frac{d^2Q}{d\tau^2}+\tau_R\frac{dQ}{d\tau}=F(Q),
\end{eqnarray}
where
\begin{subequations}
\begin{alignat}{4}
T^2&\equiv \frac{LC_sR_C}{R_L},\\
\tau_R&\equiv R_CC_s+\frac{L}{R_L}.
\end{alignat}
\end{subequations}
and
\begin{eqnarray}
\label{2}
F(Q)=\beta(Q-Q_1)(Q-Q_2),
\end{eqnarray}

We can also improve the approximation (\ref{we}) by modifying it to
\begin{eqnarray}
\label{cube}
V(Q)=\text{const}+\frac{Q+\beta Q^2+\gamma Q^3}{C_s},
\end{eqnarray}
 where $\gamma$ is also  assumed to be constant.
This modification can be important when  $Q_1,Q_2$ are close to zero and, as it normally happens, the charge is an odd function of the voltage. 
In this cases (\ref{29}) is still valid but   (\ref{2}) becomes
\begin{eqnarray}
\label{29b}
F(Q)=\gamma(Q-Q_1)(Q-Q_2)(Q+Q_3),
\end{eqnarray}
where
\begin{eqnarray}
Q_3\equiv \frac{\beta}{\gamma}+Q_1+Q_2.
\end{eqnarray}

\section{Strong dissipation}
\label{hf}

\subsection{The ODE which doesn't contain explicitly the independent variable}
\label{re}

Consider
the generalized  damped Helmholtz-Duffing  equation \cite{kogan1}
\begin{eqnarray}
\label{99}
x_{\tau\tau}+k x_{\tau}
=\gamma x\left(x^n-x_1\right)\left(x^n+x_3\right),
\end{eqnarray}
where $n$, $k$, $\gamma$,  $x_1$, $x_3$  are  constants,
with the boundary conditions
\begin{eqnarray}
\label{b}
\lim_{\tau\to-\infty}x(\tau)=x_1^{1/n},\hskip 1 cm \lim_{\tau\to+\infty}x(\tau)=0.
\end{eqnarray}

Equation (\ref{99})  doesn't contain explicitly the independent variable $\tau$. This prompts the idea to
consider $x$ as the new independent variable and
\begin{eqnarray}
\label{cot}
p=\frac{dx}{d\tau}
\end{eqnarray}
as the new dependent variable.
In the new variables
(\ref{99})  takes the form  of Abel equation of the second kind   \cite{polyanin}.
\begin{eqnarray}
\label{pp}
pp_{x}+k p=\gamma x\left(x^n-x_1\right)\left(x^n+x_3\right).
\end{eqnarray}
The boundary conditions in the new variables are
\begin{eqnarray}
\label{bmp}
p\left(x_1^{1/n}\right)= p(0)=0.
\end{eqnarray}
One can easily check up that for
$\gamma$ and $k$ connected by the formula
\begin{eqnarray}
\label{kkk}
k^2=\frac{\gamma\left[x_1+(n+1)x_3\right]^2}{n+1},
\end{eqnarray}
the solution of (\ref{pp}) satisfying the boundary conditions (\ref{bmp}) is
\begin{eqnarray}
\label{pmx}
p=mx\left(x^n-x_1\right),\hskip 1cm
m=\sqrt{\frac{\gamma}{n+1} }.
\end{eqnarray}
Substituting $p(x)$ into (\ref{cot})  and integrating  we obtain the solution of (\ref{99}) as
\begin{eqnarray}
\label{mm}
x(\tau)=\frac{x_1^{1/n}.}
{\left[\exp\left(nmx_1\tau\right)+1\right]^{1/n}}.
\end{eqnarray}

Now consider the equation
\begin{eqnarray}
\label{19}
x_{\tau\tau}+k x_{\tau}
=\beta x\left(x^n-x_1\right)
\end{eqnarray}
with the boundary conditions (\ref{b}).
We can present (\ref{19}) as
\begin{eqnarray}
\label{19b}
x_{\tau\tau}+k x_{\tau}
=\beta x\left(x^{n/2}-x_1^{1/2}\right)\left(x^{n/2}+x_1^{1/2}\right).
\end{eqnarray}
Hence we realize that for
$k$ and $\beta$ connected by the formula
\begin{eqnarray}
k=(n+4)\sqrt{\frac{\beta x_1}{2(n+2)}},
\end{eqnarray}
the solution of (\ref{19}) satisfying the boundary conditions (\ref{bmp}) is
\begin{eqnarray}
\label{mm2}
x(\tau)=\frac{x_1^{1/n}.}
{\left\{\exp\left[n\sqrt{\frac{\beta x_1}{2(n+2)}}\tau\right]+1\right\}^{2/n}}.
\end{eqnarray}

Let us return to (\ref{99})  and modify it to
\begin{eqnarray}
\label{99b}
x_{\tau\tau}+k(1+2\beta x^n) x_{\tau}
=\gamma x\left(x^n-x_1\right)\left(x^n+x_3\right).
\end{eqnarray}
Thus we take into account possible nonlinearity of friction. Thus instead of (\ref{pp}) we obtain
\begin{eqnarray}
\label{pp2b}
pp_{x}+k(1+2\beta x^n) p=\gamma x\left(x^n-x_1\right)\left(x^n+x_3\right).
\end{eqnarray}
Acting as above we obtain
that for
$k$ and $\gamma$ connected by the formula
\begin{eqnarray}
\label{kk}
k^2=\frac{\gamma [x_1+(n+1)x_3]^2}{\left(1-2\beta x_3\right)
\left(n+1+2\beta x_1\right)},
\end{eqnarray}
the solution of (\ref{pp2b}) satisfying the boundary conditions (\ref{bmp}) is
(\ref{pmx}) (same as it was for $\beta=0$), only
 this time
\begin{eqnarray}
\label{mk}
m= \sqrt{\frac{\gamma\left(1-2\beta x_3\right)}{n+1+2\beta x_1}}.
\end{eqnarray}
Hence  the solution of (\ref{99b}) with $\beta\neq 0$
is of the same form   as for $\beta=0$ (Eq. (\ref{mm})), only with the modified $m$.

\subsection{Back to the transmission line}

Now let us return to the transmission line.
For Eq. (\ref{29}) with $F(Q)$ given by (\ref{29b}),  using the results of the previous Section, we claim that when the parameters of the equation  satisfy the relation
 \begin{eqnarray}
\label{x}
\frac{\tau_R^2}{T^2}=\frac{\gamma\left(Q_1+Q_2+2Q_3\right)^2}{2},
\end{eqnarray}
the solution of  the equation  can be expressed in terms of elementary functions:
\begin{eqnarray}
\label{oib}
Q=Q_2+\frac{Q_1-Q_2}{\exp\left(\psi\tau \right)+1},
\end{eqnarray}
where
\begin{eqnarray}
\label{hrua}
\psi=\sqrt{\frac{\gamma}{2}}\cdot\frac{Q_1-Q_2}{T}.
\end{eqnarray}
For (\ref{29})  with $F(Q)$ given by (\ref{2}), we claim that when the boundary conditions  satisfy the relation
\begin{eqnarray}
\label{hru2a}
\frac{\tau_R}{T}=5\sqrt{\frac{\beta\left(Q_1-Q_2\right)}{6}},
\end{eqnarray}
the solution of the equation can be expressed in terms of elementary functions:
\begin{eqnarray}
\label{oi}
Q=Q_2+\frac{Q_1-Q_2}{\left[\exp\left(\chi\tau \right)+1\right]^2},
\end{eqnarray}
where
\begin{eqnarray}
\label{chichi}
\chi=\sqrt{\frac{\beta (Q_1-Q_2)}{6}}\cdot\frac{1}{T}.
\end{eqnarray}

We used previously the linear approximation
for $V(Q)$ in the last term in the l.h.s. of (\ref{199}). Strictly speaking, since we are considering nonlinear $V(Q)$ in the r.h.s. of (\ref{199}) more consistent would be to treat  the same way the l.h.s.
Equations. (\ref{kk}), (\ref{mk}) allow us to go one step in this direction, that is  to consider
$V(Q)$ in the last term in the l.h.s. of (\ref{199}) in quadratic approximation (for cubic nonlinearity of $V(Q)$). As the result,  Eqs. (\ref{x}) and (\ref{oib}) slightly change and  Eq. (\ref{oib}) doesn't change at all.

The analytic results for the profile of the shock wave were obtained 
 by reducing  the second order differential equation (\ref{199})  
to the first order one
\begin{eqnarray}
\label{290}
T^2p\frac{dp}{dQ}+\tau_Rp=F(Q)
\end{eqnarray}
where $p\equiv dQ/d\tau$.
However, looking back to Eq. (\ref{er}) we see the opportunity to reduce the system of two first order 
differential equations to the single first order one
\begin{eqnarray}
\label{a}
\frac{1}{R_LR_C}\frac{d \Phi}{d Q}=
\frac{U(Q-Q_2)-I(\Phi)+I_2}{U(\Phi-\Phi_2)-V(Q)+V_2}
\end{eqnarray}
already at this stage.
It would be interesting to try to obtain exact analytic solutions of (\ref{a}) for some values of the parameters.

Postponing such attempt until later time, note here that Eq. (\ref{a}) allows us to obtain the $\tau$ dependence of the current $I$ which was absent in our previous publications. To achieve this aim let us rewrite (\ref{a}) in the form
\begin{eqnarray}
\label{a7}
I=I_2+U(Q-Q_2)\nonumber\\
-\frac{L}{R_LR_C}\left[UL(I-I_2)-V(Q)+V_2\right]\frac{d I}{d Q}.
\end{eqnarray}
The dependence $Q(\tau)$ being found earlier, Eq. (\ref{a7}) gives infinite series for $I(\tau)$, and the first two terms of the series are
\begin{eqnarray}
\label{ii}
I=I_2+U(Q-Q_2)-\frac{L}{C_sR_LR_C} \frac{I_2-I_1}{Q_2-Q_1}F(Q).
\end{eqnarray}
Thus from (\ref{oib}) follows
\begin{eqnarray}
I=I_2+\frac{I_1-I_2}{\exp\left(\psi\tau \right)+1}
\left\{1+\frac{L(Q_1-Q_2)}{C_sR_LR_C}
\right.\nonumber\\
\left.\frac{\exp\left(\psi\tau \right)}{\exp\left(\psi\tau \right)+1} \left[\frac{Q_1-Q_2}{\exp\left(\psi\tau \right)+1}
+\frac{\beta}{\gamma}+Q_1+2Q_2\right]\right\},
\end{eqnarray}
and from(\ref{oi}) follows
\begin{eqnarray}
I=I_2+\frac{I_1-I_2}{\left[\exp\left(\chi\tau \right)+1\right]^2}
\left\{1+\frac{L(Q_1-Q_2)}{C_sR_LR_C}\right.\nonumber\\
\left.\left[1-\frac{1}{\left[\exp\left(\chi\tau \right)+1\right]^2}\right]\right\}.
\end{eqnarray}

\section{Weak dissipation}

Let us return to Eq. (\ref{29}) and consider the case of weak dissipation.
In this case it is convenient to rewrite the equation as 
\begin{eqnarray}
\label{dvec}
\frac{d^2Q}{d\widetilde{\tau}^2}+\frac{d\Pi(Q)}{dQ}=-\kappa\frac{dQ}{d\widetilde{\tau}},
\end{eqnarray}
where $\widetilde{\tau}\equiv\tau/T$, $\kappa\equiv\tau_R/T$, and $\Pi(Q)$ is defined by the equation
\begin{eqnarray}
\frac{d\Pi(Q)}{dQ}\equiv F(Q).
\end{eqnarray}
We will use  the method of time averaging \cite{kog}.
We assume that we know the undamped solution
 $Q_{ud}(\tau;{\cal E})$,  satisfying equation
\begin{eqnarray}
\label{con}
\frac{1}{2}\left(\frac{dQ_{ud}}{d\widetilde{\tau}}\right)^2+\Pi(Q_{ud})={\cal E}\,,
\end{eqnarray}
and express    damped oscillations as
\begin{eqnarray}
\label{time}
Q(\widetilde{\tau})=Q_{ud}(\widetilde{\tau};{\cal E}(\widetilde{\tau})),
\end{eqnarray}
where ${\cal E}(\widetilde{\tau})$ satisfies equation
\begin{eqnarray}
\label{eee}
\frac{d {\cal E}}{d\widetilde{\tau}}=-\kappa\left<\left(\frac{d Q_{ud}}{d\widetilde{\tau}}\right)^2\right>_{\cal E};
\end{eqnarray}
the averaging is with respect to the period of the undamped oscillation with the energy ${\cal E}$, or more explicitly, 
\begin{eqnarray}
\label{eeey}
\frac{d {\cal E}}{d\widetilde{\tau}}=-
\frac{2\kappa\int dQ\sqrt{{\cal E}-\Pi(Q)}}{\int dQ/\sqrt{{\cal E}-\Pi(Q)}};
\end{eqnarray}
the limits of integration in both integrals are found from the equation
${\cal E}-\Pi(Q)=0$.
Notice that the shock in the considered case can be called the stationary dispersive shock wave \cite{kamchatnov}.  

We want to show how the method of time averaging works for $F(Q)$ given by Eq. (\ref{2}). After the change of  variable
\begin{eqnarray}
\label{tr}
\widetilde{Q}=\frac{\beta}{6}\left(Q-\frac{Q_1+Q_2}{2}\right),
\end{eqnarray}
Eq. (\ref{con}) takes the form
\begin{eqnarray}
\label{w34}
\left(\frac{d\widetilde{Q}_{ud}}{d\widetilde{\tau}}\right)^2=4\widetilde{Q}_{ud}^3
-g_2\widetilde{Q}_{ud}-g_3,
\end{eqnarray}
where
\begin{subequations}
\begin{alignat}{4}
g_2&=\beta^2\frac{\left(Q_1-Q_2\right)^2}{12}\\
g_3&=-\frac{\beta^2}{18}{\cal E}.
\end{alignat}
\end{subequations}
Equation (\ref{w34}) defines Weierstrass elliptic function \cite{abram}.
\begin{eqnarray}
\label{gy}
\widetilde{Q}(\widetilde{\tau};{\cal E})={\cal P}(\widetilde{\tau};g_2,g_3)\,.
\end{eqnarray}
Thus the  solution of (\ref{dvec}) is
\begin{eqnarray}
\label{psih}
Q(\widetilde{\tau})={\cal P}(\widetilde{\tau};g_2,g_3(\widetilde{\tau}))\,.
\end{eqnarray}

Equation (\ref{eeey}) can be rewritten as
\begin{eqnarray}
\frac{d g_3}{d\widetilde{\tau}}=4\kappa\frac{{\cal N}(g_3)}{Y_0(g_3)},
\end{eqnarray}
where
\begin{subequations}
\label{ellip}
\begin{alignat}{4}
Y_0(g_3)=\int_c^b \frac{d\psi}{\sqrt{P(\psi)}}\,&,\;\;\;
{\cal N}(g_3)=\int_c^b d\psi \sqrt{P(\psi)},\label{ellipa}\\
P(\psi)&=4\psi^3-g_2\psi-g_3;
\end{alignat}
\end{subequations}
 $a,b,c$ ($a>b>c$)
 are  the  roots of  cubic  equation $P(\psi)=0$.
 
All integrals
\begin{eqnarray}
Y_m=\int\frac{\psi^m d\psi}{\sqrt{P(\psi)}}\,,
\end{eqnarray}
where $m$ is an arbitrary natural number and $P(\psi)$ is some polynomial of power $p$,
are expressed through the  $p-1$ first integrals $Y_0,Y_1,\dots,Y_{p-2}$ and algebraic quantities \cite{goursat}. 
$Y_0$ and $Y_1$ are table integrals \cite{grad}:
\begin{subequations}
\label{gou}
\begin{alignat}{4}
Y_0&=\frac{2}{\sqrt{a-c}}K(k) \, ,\label{gou1}\\
Y_1&=\frac{2a}{\sqrt{a-c}}K(k)-2\sqrt{a-c}E(k) \,,\label{gou2}
\end{alignat}
\end{subequations}
where $K$ and $E$ are complete elliptic integrals of the first  and second kind respectively, and $k=\sqrt{(b-c)/(a-c)}$.
To calculate ${\cal N}(g_3)$ let us integrate the identity
\begin{eqnarray}
\sqrt{P(\psi)}=\frac{4\psi^3-g_2\psi-g_3}{\sqrt{P(\psi)}}.
\end{eqnarray}
We obtain
\begin{eqnarray}
\label{gour}
{\cal N}=4Y_3-g_2Y_1-g_3Y_0\,.
\end{eqnarray}
Integrating the identity
\begin{eqnarray}
\frac{d}{d\psi}\left[\psi\sqrt{P(\psi)}\right]=\sqrt{P(\psi)}
+\frac{12\psi^3-g_2\psi}{2\sqrt{P(\psi)}},
\end{eqnarray}
we obtain
\begin{eqnarray}
\label{gourd}
2{\cal N}+12Y_3-g_2Y_1=0\,.
\end{eqnarray}
Combining (\ref{gour})   and (\ref{gourd}) we obtain
\begin{eqnarray}
\label{gg}
{\cal N}=-\frac{1}{5}\left(2g_2Y_1+3g_3Y_0 \right)
\end{eqnarray}
and, finally,
\begin{eqnarray}
\label{ou}
\frac{d g_3}{d\widetilde{\tau}}=\frac{4\kappa}{5}
\left[ag_2+3g_3-(a-c)g_2\frac{E(k)}{K(k)}\right]\,.
\end{eqnarray}
(One should keep in mind that $a,c,k$ are functions of $g_3$.)
Numerical analysis of the formula obtained in this Section one may find in our previous publication \cite{kog}.

\section{The simple waves in a lossless nonlinear transmission line}

As it is well known \cite{landau}, the shock waves exist only in the presence of dissipation. (In the absence of the dissipation, the localized travelling waves in the systems we consider are kinks and solitons \cite{kogan1}.) This corresponds to the case of a 
lossy nonlinear transmission line, considered in the previous part of the paper.

\begin{figure}[h]
\includegraphics[width=\columnwidth]{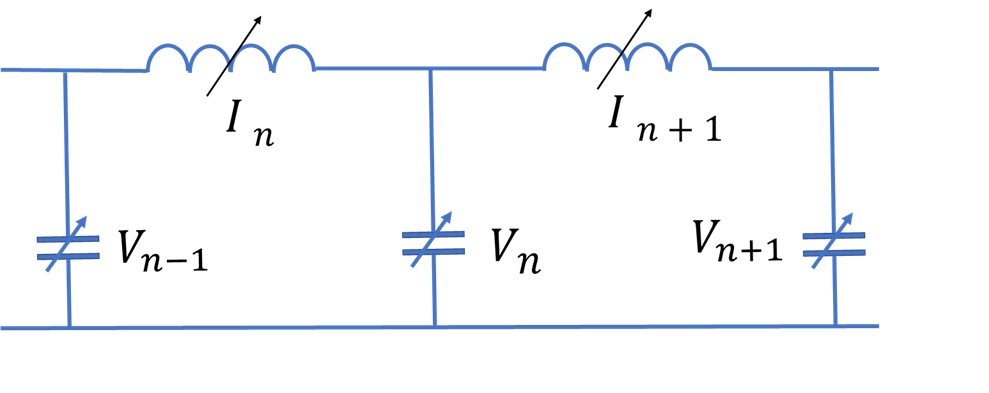}
\caption{Lossless nonlinear  transmission line}
 \label{trans5}
\end{figure}

On the other hand, many properties of the shocks in fluids were studied beginning from the XIX century without taking the dissipation explicitly into account, but just postulating the existence of discontinuous solutions and the jump conditions \cite{krehl}.
Similarly to that, the initial phase of the formation of a shock wave can be considered ignoring the dissipation.
The present section is dedicated to a lossless   transmission line, constructed from the identical nonlinear inductors and the identical nonlinear capacitors as shown on Fig. \ref{trans5}.
We take the  capacitors voltages $V_n$ and the currents through the inductors
$I_n$
as the dynamical variables.
The circuit equations (Kirchhoff laws) are
\begin{subequations}
\label{a10}
\begin{alignat}{4}
C(V_n)\frac{dV_n}{dt}&=  I_n-I_{n+1},\label{a8b}\\
L(I_{n+1})\frac{d I_{n+1}}{d t}&=V_{n}-V_{n+1}.\label{8a}
\end{alignat}
\end{subequations}
Further on we'll consider $C(V)$ and $L(I)$ as known functions.

In the continuum approximation we  treat $n$  as the continuous variable $x$
(we measure distance in the units of the transmission line period)
and approximate the finite differences in the r.h.s. of the equations by the first derivatives with respect to $x$,   after which the equations take the form
\begin{subequations}
\label{fe}
\begin{alignat}{4}
C(V)\frac{\partial V}{\partial t} &=  -\frac{\partial I}{\partial x},\\
L(I)\frac{\partial I}{\partial t}&= -\frac{\partial V}{\partial x}.
\end{alignat}
\end{subequations}

The simple wave approximation allows  to decouple the wave equations into two separate equations for the right- and left-going waves \cite{landau,rabinovich,vinogradova}.
In our previous publications we introduced such approximation for Eq. (\ref{fe}) for the case of half-nonlinear transmission line (more specifically for the case of JTL \cite{kogan1}).
In the present Section we formulate the simple wave approximation for the general lossless nonlinear transmission line including both the nonlinear inductors and the nonlinear capacitors.

The simple wave approximation, well known in the theory of nonlinear waves, refers to a wave where the relationship between the wave's properties (like displacement, pressure, or density) and its spatial and temporal coordinates can be described by a single, independent function. In essence, it's a wave where the relationship between its amplitude and other characteristics is straightforward, and not influenced by other waves or complex interactions. More specifically, it allows to decouple the waves moving in the opposite
directions (in a 1D case).

To formulate such approximation in our case, let us start from the small amplitude  waves
on a homogeneous background $ V_0,I_0$.
\begin{subequations}
\begin{alignat}{4}
V&= V_0+v,\\
I&=I_0+i .\label{i}
\end{alignat}
\end{subequations}
For such waves Eq. (\ref{fe}) is simplified to
\begin{subequations}
\label{ve9d}
\begin{alignat}{4}
C(Q)\frac{d v}{d t} &=  -\frac{di}{d x},\\
L(I)\frac{d i}{d t}&= -\frac{d v}{d x},
\end{alignat}
\end{subequations}
(for brevity we have discarded lower index 0 in Eq. (\ref{ve9d})).

The  solutions of Eq. (\ref{ve9d})  are right- and left-propagating travelling waves, each   depending upon the single variable
$\tau_{\pm}=t\mp x/u_0$, propagating with the speed 
\begin{eqnarray}
 U(I,V)=\frac{1}{\sqrt{L(I) C(V)}}. 
\end{eqnarray}
The voltage and current  in the "sound" wave are
 connected by the equation
\begin{eqnarray}
v=\pm Z(I,V)i,
\end{eqnarray}
where
\begin{eqnarray}
 Z(I,V)=\sqrt{\frac{L(I)}{C(Q)}}.
\end{eqnarray}

The simple wave approximation, that is  decoupling of (\ref{fe}) into two separate equations for the right- and left-going waves,
is achieve by considering $V$ as a function of $I$ (or vice-verse). 
We emphasize that our approximation (see below) is based on previously known approach but the application of the approximation to the fully nonlinear transmission line was (to the best of our knowledge) was never done before.

 Then from Eq. (\ref{fe}) we obtain
\begin{eqnarray}
\frac{C(V)}{L(I)}\frac{d V}{d I} =\frac{d I}{d V}, 
\end{eqnarray}
or equivalently
\begin{eqnarray}
\frac{d V}{d I} =\pm Z(I,V).
\end{eqnarray}
Substituting into (\ref{fe}) we obtain a system of two coupled equations for each of the simple waves
\begin{subequations}
\label{fe2}
\begin{alignat}{4}
\frac{\partial I}{\partial t} &= \pm U(I,V)\frac{\partial I}{\partial x},\label{fea}\\
\frac{\partial V}{\partial t}&= \pm U(I,V)\frac{\partial V}{\partial x}. \label{feb}
\end{alignat}
\end{subequations}
Further on for the sake of definiteness we'll talk only about the right-going wave which corresponds to taking the sign minus in the r.h.s. of both equations in (\ref{fe2}).

The system (\ref{fe2}) simplifies in  half-nonlinear cases, that is when either the capacitor or the inductor is linear. In the first case ($C(V)=$const)
\begin{eqnarray}
U=U(I),
\end{eqnarray}
and (\ref{fea}) becomes closed equation for the current
\begin{eqnarray}
\label{1}
\frac{\partial I}{\partial t} + U(I)\frac{\partial I}{\partial x}=0.
\end{eqnarray}
Instead of (\ref{feb}) we can use equation
\begin{eqnarray}
\label{3}
U=Z(I)I.
\end{eqnarray}
Talking about this case we have in mind first and foremost the Josephson transmission line. Both Josephson laws can be presented as 
\begin{eqnarray}
L(I)=\frac{\hbar}{2e\sqrt{I_c^2-I^2}},
\end{eqnarray}
thus we obtain 
\begin{eqnarray}
U(I)=\sqrt{\frac{2e}{\hbar C}}\sqrt[4]{I_c^2-I^2}.
\end{eqnarray}

However, the fully nonlinear case can also be treated easily.
If we consider the initial value problem
\begin{eqnarray}
I(x,0)=I_0(x),\hskip 1 cm V(x,0)=V_0(x),
\end{eqnarray}
the solution of Eq. (\ref{fe2}) can be obtained by inspection \cite{logan}
\begin{subequations}
\label{fe23}
\begin{alignat}{4}
I(x,t) &= I_0(\xi)\\
V(x,t) &= V_0(\xi),
\end{alignat}
\end{subequations}
where
\begin{eqnarray}
x-\xi=U(V_0(\xi),I_0(\xi))t.
\end{eqnarray}

The simple wave approximation (\ref{fe2}) can be applied also to Eq. (\ref{c2}). In fact, the dissipative terms in the equation determine the profile of the shock. On the other hand, if we want to study the formation of the shocks, then assuming these terms to be in some sense small, we can ignore the influence of the dissipation on the process of the formation (until we don't approach to close to the singularity of the dissipationless equation).
Thus ignoring the dissipation in (\ref{c2}) we may rewrite the equation in the form
\begin{subequations}
\label{feu}
\begin{alignat}{4}
\frac{dQ}{dV}\frac{\partial V}{\partial t} &=  -\frac{\partial I}{\partial x},\\
\frac{d\Phi}{dI}\frac{\partial I}{\partial t}&= -\frac{\partial V}{\partial x},
\end{alignat}
\end{subequations}
which coincides with Eq. (\ref{fe}) if we put
\begin{eqnarray}
C(V)=\frac{dQ}{dV},\hskip 1cm L(I)=\frac{d\Phi}{dI}.
\end{eqnarray}
After that we can apply the procedure presented above in this Section.

Now let us forget about the dissipation and consider the strictly disssipationless case.
If we want to study the profile of the travelling waved in such case, additional complication
(with respect to the lossy case) arises. 
In the latter case
we started from considering the discrete transmission line, but the presence of the dissipation introduced the space scale into the system, and this scale was implicitly assumed to be much larger than the period of the line. This allowed us to use 
the continuum approximation, actually ignoring the discrete nature of the system.
For the lossless case the scale of the localized travelling wave  is determined by the period of the transmission line \cite{kogan1}. This makes the continuum approximation inadequate and we introduced the quasi-continuum approximation \cite{kogan1}, 
which corresponds to 
approximating the finite differences in the r.h.s. of the equations (\ref{a10}) by the two first terms in the Taylor expansion \cite{kogan1}. Thus instead of Eq. (\ref{fe})
we obtain
\begin{subequations}
\label{ve9c}
\begin{alignat}{4}
C(Q)\frac{\partial V}{\partial t} &=  -\frac{\partial I}{\partial x}
-\frac{1}{24}\frac{\partial^3 I}{\partial x^3},\label{vb}\\
L(I)\frac{\partial I}{\partial t}&= -\frac{\partial V}{\partial x}
-\frac{1}{ 24}\frac{\partial^3 V}{\partial x^3}. \label{vb0}
\end{alignat}
\end{subequations}

We studied in details the travelling waves described by these equations for the case of half-nonlinear transmission line (more specifically for the case of JTL \cite{kogan1}).
In distinction to the lossy case, where the travelling waves turn out to be the shocks, 
in the lossless case the travelling waves turn out to be the kinks and the solitons.

We also studied (with much less details) the formation of the solitons and the kinks via introducing the simple wave approximation for the JTL. Now we want
to formulate the simple wave approximation for the general lossless nonlinear transmission line including both the nonlinear inductors and the nonlinear capacitors on top of the quasi-linear approximation.

One must understand that Eq. (\ref{fe}) (and hence Eq. (\ref{fe2}) can describe the formation of the kinks and the solitons
(until we don't approach to close to the singularities of the equations).
 Our present aim is to formulate the approximation which will describe both the formation of the kinks and the solitons and their profiles.

Starting from Eq. (\ref{ve9c}) and repeating the process which led from (\ref{fe}) to (\ref{fe2}) we obtain instead of the latter
\begin{subequations}
\label{f}
\begin{alignat}{4}
\frac{\partial I}{\partial t} &= \pm U(I,V)\left(\frac{\partial I}{\partial x}
+\frac{1}{24}\frac{\partial^3 I}{\partial x^3}\right),\label{fa}\\
\frac{\partial V}{\partial t}&= \pm U(I,V)\left(\frac{\partial V}{\partial x}
+\frac{1}{24}\frac{\partial^3 V}{\partial x^3}\right). \label{fb}
\end{alignat}
\end{subequations}

Let us apply thus improved simple wave approximation to the JTL. In this case instead of Eq. (\ref{1}) we obtain
\begin{eqnarray}
\label{om}
\frac{\partial I}{\partial t}
+\sqrt{\frac{2e}{\hbar C}}\sqrt[4]{I_c^2-I^2}\left(\frac{\partial I}{\partial x}
+\frac{1}{24}\frac{\partial^3 I}{\partial x^3}\right)=0.
\end{eqnarray}
If we make an additional assumption $|I|\ll I_c$, Eq. (\ref{om}) can be written down as \cite{kogan1}
\begin{eqnarray}
\label{omm}
\frac{\partial I}{\partial t}
+\sqrt{\frac{2eI_c}{\hbar C}}\left(\frac{\partial I}{\partial x}-
\frac{1}{12I_c^2}\frac{\partial I^3}{\partial x}
+\frac{1}{24}\frac{\partial^3 I}{\partial x^3}\right)=0,
\end{eqnarray}
where we have ignored the term proportional to $I^2\partial^3 I/\partial x^3$.
Looking at Eq. (\ref{omm}) we recognize 
the modified Korteweg-de Vries (mKdV) equation  \cite{drazin}.

On the other hand, considering small variations of the current on the constant background
presented by Eq. (\ref{i}), from (\ref{omm}) we obtain \cite{kogan1}
\begin{eqnarray}
\label{om4}
\frac{\partial i}{\partial t}
+\sqrt{\frac{2e}{\hbar C}}\sqrt[4]{I_c^2-I_0^2}\nonumber\\\left(\frac{\partial i}{\partial x}
-\frac{I_0}{4\left(I_c^2-I_0^2\right)}\frac{\partial i^2}{\partial x}
+\frac{1}{24}\frac{\partial^3 i}{\partial x^3}\right)
=0,
\end{eqnarray}
where we have ignored the term proportional to $i\partial^3 i/\partial x^3$.
Looking at Eq. (\ref{om4}) we recognize 
the  Korteweg-de Vries (KdV) equation  \cite{drazin}.

To conclude we state that we obtained the exact analytical expressions for the profile of the shock waves (both the current and the voltage)  in half-nonlinear transmission lines for the appropriate values of the parameters. We also formulated the simple wave approximation for the lossless discrete nonlinear  transmission line.

\end{document}